\pdfoutput=1
\documentclass[prl,aps,showpacs,showkeys,amsmath,amssymb,twocolumn]{revtex4}
\usepackage{bm}
\usepackage[english]{babel}
\usepackage{graphicx}

\newcommand{\sech}{\text{sech}}

\renewcommand{\Re}{\text{Re}}

\begin{document}
\author{Kirill Rivkin}
\affiliation{Seagate Technology, Bloomington, Minnesota 55435}
\author{Konstantin Romanov}
\email{const@physics.tamu.edu}
\author{Yury Adamov}
\author{Artem Abanov}
\author{Valery Pokrovsky}
\author{Wayne Saslow}
\affiliation{Department of Physics, Texas AM University, College Station, TX, 77843}
\title{Continuous N\'eel to Bloch Transition as Thickness Increases: Statics and Dynamics}

\begin{abstract}
We analyze the properties of N\'eel and Bloch domain walls as a 
function of film thickness $h$,
for systems where, in addition to exchange, the dipole-dipole interaction must be included.
The N\'eel to Bloch phase transition is found to be a second order
transition at $h_{c}$, mediated by a single unstable mode that corresponds to 
oscillatory motion of
the domain wall center.  A uniform out-of-plane rf-field 
couples strongly to this critical mode only in the N\'eel phase.
An analytical Landau theory 
shows that the critical mode frequency $\omega \sim \sqrt{h_c -h}$
just below the transition, as found numerically.
\end{abstract}


\maketitle

{\bf Introduction.} Domain walls in ferromagnetic thin films, 
including exchange, uniaxial anisotropy, and 
the dipole-dipole interaction, have been extensively 
studied for the past 100 years.\cite{bitter:1903}
For a monodomain of small film thickness $h$, the dipole-dipole 
interaction causes the magnetization to lie entirely in the plane. 
Such domains in a thin film are separated by 
either Bloch or N\'eel domain wall. 
For the Bloch domain wall 
the transition 
between domains occurs with the magnetization developing an 
out-of-plane component.\cite{bloch:1932,landau:1935} 
The system develops surface poles, 
and the associated total dipole-dipole energy is proportional 
to $h$. For the N\'eel domain wall 
the magnetization lies entirely in the
plane of the film.\cite{neel:1955,middlehoek:1966} 
The total dipole-dipole energy for a N\'eel wall comes from 
the self-interaction of a volume pole density, and thus is 
proportional to $h^{2}$.
It is thus clear that for small thickness $h$ the N\'eel wall is energetically
preferable, whereas at larger $h$ the Bloch wall wins. 
This N\'eel-Bloch domain wall transition is the major focus of this letter. 

Recent experimental and micromagnetic studies show that a Bloch wall has a 
complex structure.\cite{scheinfein:1989,scheinfein:1991}  Its interior magnetization has an 
out-of-plane (Bloch-like) component, whereas its near-surface 
magnetization is in-plane (N\'eel-like) --- the so-called N\'eel caps (Figure 1). 

The N\'eel domain wall, on the other hand, has a narrow central part which 
resembles the usual ``exchange'' domain wall and long logarithmic 
tails\cite{melcher:2003,RivkinRomanov-Neel:2007}. These tails are the consequence of the
dipole-dipole interaction.

When the sample is sufficiently wide the so-called cross-tie domain wall 
can be an alternative equilibrium configuration.\cite{chang:2007} 
In this letter we completely neglect such configurations.
We assume that at the thicknesses of interest 
the system possesses a single domain wall. For $h$ less than the 
critical thickness $h_{c}$, the domain wall is of N\'eel type, 
whereas for $h>h_{c}$ the magnetization in the domain wall itself 
tips out-of-plane, forming so-called symmetric and asymmetric 
Bloch walls.\cite{redjdal:2002} For the thicknesses considered here, 
N\'eel walls and asymmetric Bloch walls are the only low energy states. 

\begin{figure}[htbp]
\includegraphics[angle=-90,width=3in]{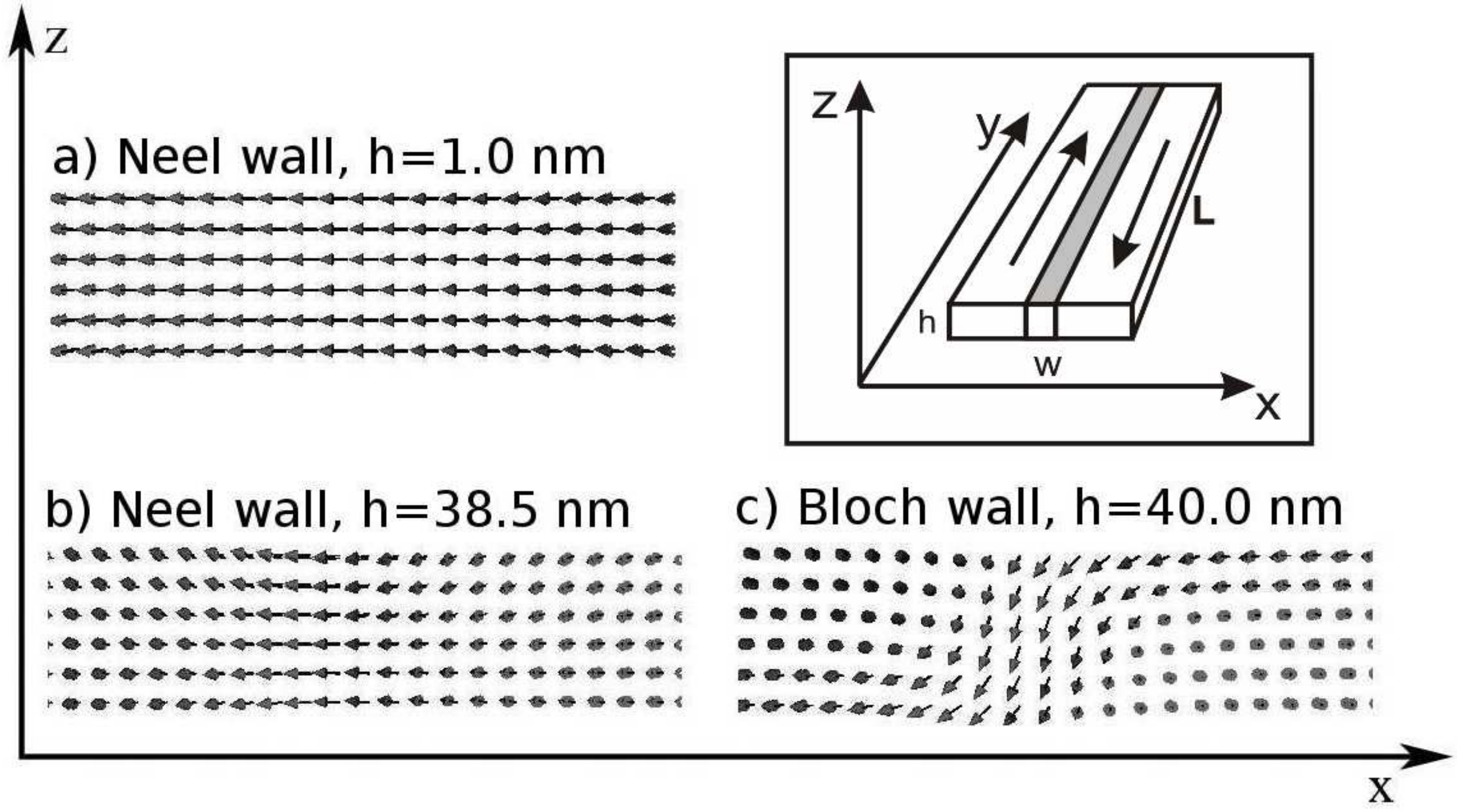}
	\vskip-0.1in
	\caption{N\'eel and asymmetric Bloch domain walls, magnetization distribution in the $x$-$z$ plane. Sample (permalloy) is infinite along $y$, has thickness $h$ along $z$ and has width $w=200$ nm along $x$.}
   \label{fig:example1}
\vskip-0.10in
\end{figure}

Despite its fundamental nature, 
it has been unclear whether the N\'eel to Bloch transition at $h_{c}$ 
is continuous or discontinuous.\cite{hubert:1975,metlov:2005,trunk:2001}  
A recent and comprehensive magnetostatics study by Kakay and 
Humphrey\cite{kakay:2005} indicates that the transition 
between N\'eel walls and asymmetric Bloch walls is first order, 
although the authors were cautious in identifying the specific thickness at 
which the transition occurs.

The present Letter considers the nature of this 
transition (continuous or discontinuous) by extending our 
work on the statics\cite{RivkinRomanov-Neel:2007} and dynamics 
for N\'eel and Bloch walls.
We numerically study\cite{rivkin:2004} the spectrum 
of normal modes as a function of the sample thickness $h$, and
identify an unstable mode whose frequency vanishes at a critical thickness $h_{c}$. 
We numerically show that the ground state of samples with $h<h_c$ is a N\'eel domain wall 
and the ground state of samples with $h>h_c$ is an asymmetric Bloch wall.

We also show that the transition can be described 
analytically by a Landau theory of second order
phase transitions with a single order parameter --- the amplitude of the 
unstable mode.
We derive the Landau free energy  for the transition up to the fourth order in
the order parameter. In agreement with the numerics, the frequency of the critical mode 
$\omega \sim \sqrt{h_{c}-h}$ just below the critical thickness.

{\bf Micromagnetics.} Consider a ferromagnetic strip of thickness $h$ along $z$, 
width $2 w$ along $x$, and infinite along $y$ (inset to Figure 1).  We assume that 
the exchange length $l_{ex}=\sqrt{A/2\pi M_{s}^{2}}$ satisfies $l_{ex}\ll w$.  
Our samples are large enough for the domain wall center to completely fit within the sample.
The material parameters chosen are appropriate to  
permalloy: exchange constant $A=1.30\times 10^{-6}$ erg/cm and saturation magnetization 
$M_s=795$ emu/cm$^3$, which gives $l_{ex}=5.72$ nm. We take no crystalline anisotropy. 
The magnetization for $x=0$ is taken to be parallel to $x$-axis and free boundary conditions
are taken on the sides of the strip. 

The calculation of the spectrum for samples with different thicknesses started from
thin samples $h$ (1--5 nm) and magnetization $\vec{m}_i^{0}=\hat{\vec{y}} M_s$, $x<0$, 
$\vec{m}_i^{0}=-\hat{\vec{y}} M_s$, $x\ge 0$.
We equilibrated and then calculated the normal modes and their coupling to a uniform rf external 
magnetic field.  Using this equilibrium as a starting point in the relaxation algorithm, 
we then gradually increased $h$ and calculated the new equilibrium configuration, normal modes, 
and rf couplings.  We also started with large thicknesses and gradually decreased the thickness, 
to check for bi-stable solutions. 
\begin{figure}[htbp]
\includegraphics[width=3.3in]{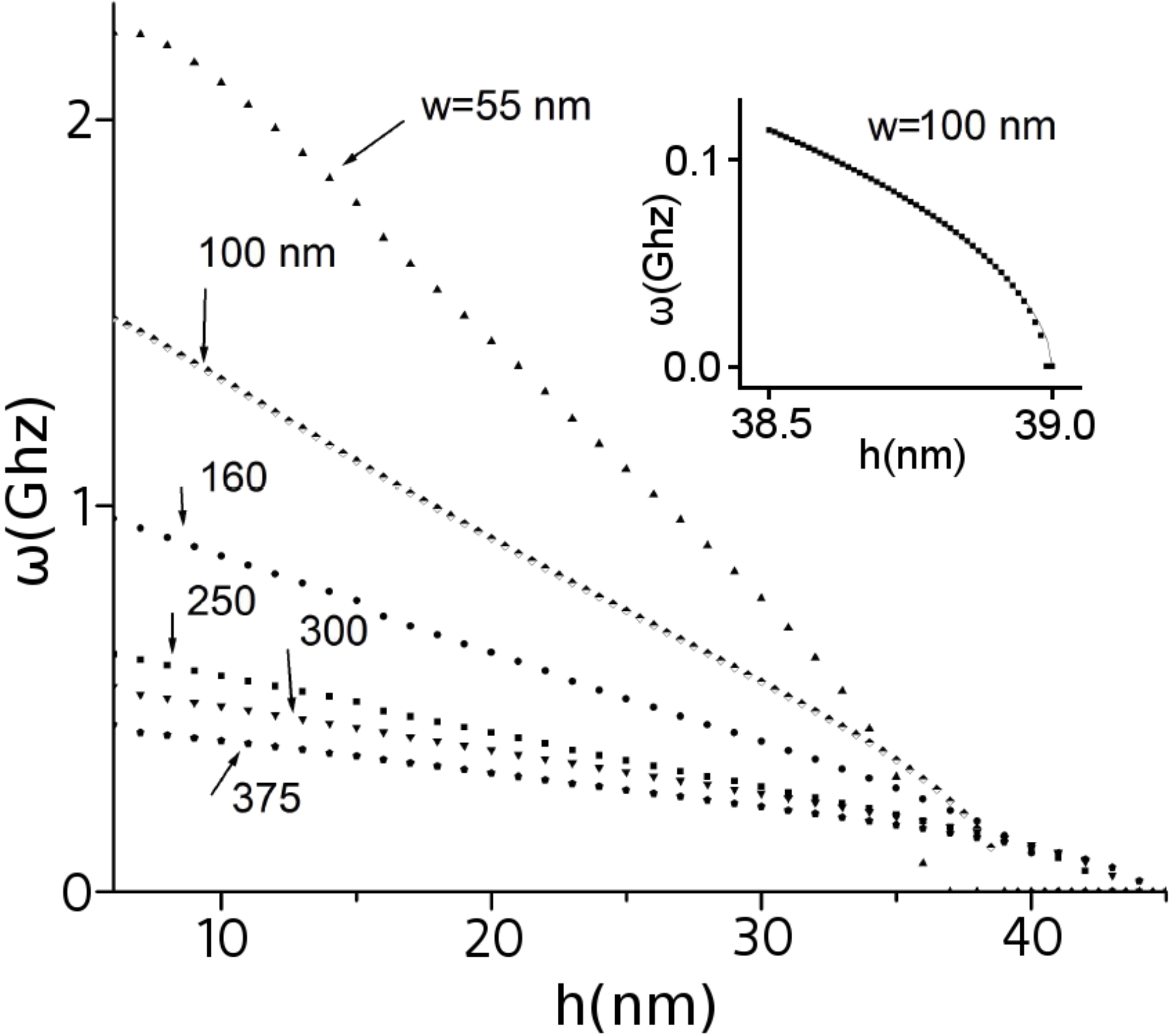}
	\caption{Frequency $\omega$ of the critical mode as a function of sample thickness $h$ for different half widths $w$. Inset: critical mode frequency near the critical thickness $h_{c}$, numerical results (dots) and square root fit (line) for $w=100$ nm.}
   \label{fig:example2}
\end{figure}

The main two results of the micromagnetics calculations are: (1) 
The ground magnetization state for small thicknesses 
is indeed the N\'eel domain wall. In this system the N\'eel wall has three 
distinct regions --- the central region of width $2\delta $ 
and logarithmic tails.\cite{RivkinRomanov-Neel:2007} 
(2) There is one mode of oscillations with frequency $\omega $ which 
goes to $0$ at some critical thickness $h_{c}$ (Fig 2). 
This mode approximately corresponds to a $z$-dependent oscillation 
of the domain wall core along $x$. The oscillations are confined to the 
central part of the N\'eel domain wall, whereas the logarithmic 
tails remain unperturbed. The frequency $\omega$ depends on both the 
width $2w$ and on the height $h$.

To a very good approximation, the (un-normalized) critical mode 
has in-plane oscillations of the form 
\begin{eqnarray}
\label{mode_x_magn_approx}
m_x^{(1)}(x,z)&\approx& \Re\left[\tanh\left(\frac{x}{\delta}\right)
\sech\left(\frac{x}{\delta}\right)
e^{i\phi(x,z)-i\omega t}\right],\\
\label{mode_y_magn_approx}
m_y^{(1)}(x,z)&\approx& \Re\left[\sech^2\left(\frac{x}{\delta}\right) 
e^{i\phi(x,z)-i\omega t}\right].
\end{eqnarray}
Here $\delta$ -- the half-width of the domain wall center -- was 
determined from a fit to the central region of the domain wall.  
The $x$-$y$ phase factor $\phi(x,z)$ determines the symmetry properties 
of the mode, but is not needed for the present purpose.  

The out-of plane oscillations $m_z^{(1)}(x,z)$ have the form
\begin{equation}
\label{mode_z_magn_approx}
m_z^{(1)}\approx \Re\left[\sech\left(\frac{x}{\delta}\right) \left( 2 \sech\left(\frac{x}{\delta}\right) -1\right)
e^{i\phi_z(x,z)-i\omega t}\right],
\end{equation}
where the $z$ phase factor $\phi_z(x,z)$ differs from $\phi(x,z)$.
The amplitude of the out-of-plane oscillations is typically 
less than 25\% of the amplitude of the 
in-plane oscillations, except for $h$ near $h_{c}$. 

\begin{figure}[htbp]
\includegraphics[width=3in]{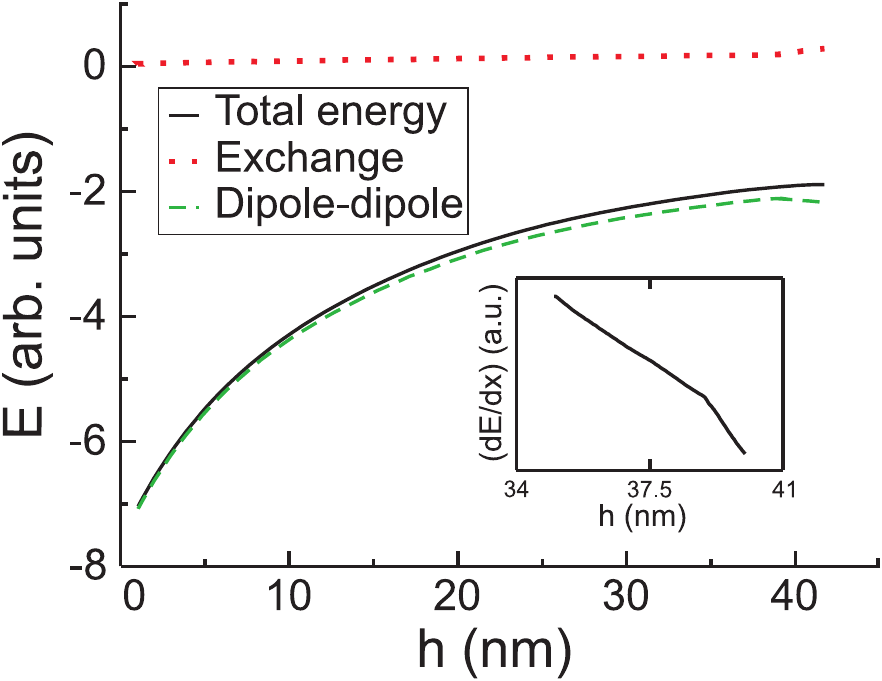}
	\caption{Magnetostatic energy and its first derivative as a function of thickness $h$ ($w=100$ nm).}
   \label{fig:example3}
\end{figure}

The inset to Figure 2 shows, for $w$ = 100 nm, the thickness dependence of $\omega$ just below the critical thickness $h_c$ (dots) as well as a fit (line) confirming the $\sqrt{h_c-h}$ dependence.  As $h$ increases above $h_{c}$, and ground state is taken as N\'eel wall, the imaginary part of $\omega$ goes from negative (stable) to positive (unstable). This is the only unstable mode.
 
Near $h_c$ conventional methods (conjugate gradient algorithm, Landau-Lifshitz equation in the presence of large damping, etc.) failed to find the equilibrium state.  This is because among all the modes in the N\'eel state only one is unstable; therefore any random perturbation of the initial configuration will contain only a negligible projection on the unstable mode. Moreover, the unstable imaginary part of the eigenfrequency is of order 1000 Hz, which would require an enormous time interval for the instability to be observed by numerical integration. The only method we found to produce reliable and consistent results was to study the normal modes for the unstable initial configuration, find the unstable mode, and then add a component of an unstable eigenvector to the static solution. This led to a significant decrease of energy and gave a magnetization close enough to the true local equilibrium that more conventional methods then applied. 

For $w=100$ nm the domain wall for $h>h_c$, where $h_c\approx39$ nm, is indeed an asymmetric Bloch wall (Figure 1). Figure 3 shows that to numerical accuracy both the total magnetostatic energy and its 
derivative remains continuous through the transition, although the dipole-dipole and exchange energies 
individually have discontinuous first derivatives. The second derivative of the magnetostatic energy 
is slightly discontinuous at the transition, which implies a second order transition. The difference between this and previous work \cite{trunk:2001} may be due to the failure of conventional numerical methods at the transition.  If the calculations are performed with a false equilibrium (metastability), then as the film gets thicker more modes become unstable.  At some thickness the numerical method finally finds the true equilibrium; at this point the transition would appear to be sudden.

We also numerically studied the dependence of $h_c$ on the 
exchange length $l_{ex}=\sqrt{A/2\pi M_s^2}$ 
and the sample width $w$. By varying these parameters and later fitting 
the results to about 5\% we find the best fit to be given by the form (Figure 4):
\begin{equation}
\label{hc_approx}
h_c=5.4 l_{ex}^{0.914} w^{0.086}
\mbox{.}
\end{equation}
\begin{figure}[htbp]
	\vskip-2.5in
\includegraphics[width=4.0in]{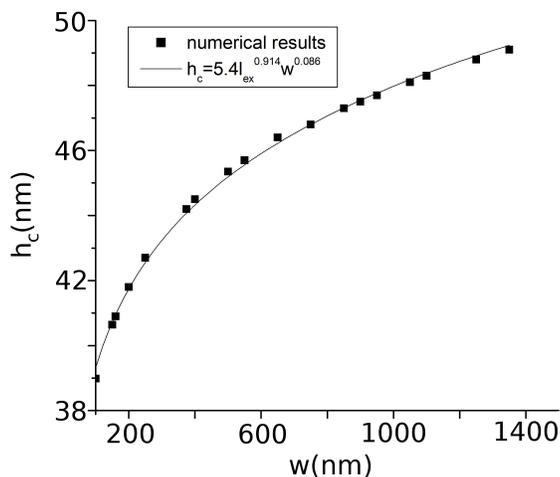}
	\vskip-0.2in
	\caption{Critical thickness $h_{c}$ as function of width $w$: dots give numerical results; curve gives fit with (\ref{hc_approx}).}   
  \label{fig:example4}
\end{figure}

{\bf Analytics.} 
This section studies the symmetry properties of the transition mode 
and its corresponding state.  It then relates the width $\delta$ of the N\'eel 
domain wall center and the thickness of the strip $h$ at the transition.

For very thin strips ($w\gg h$) below the critical thickness $h_{c}$ 
the stable magnetization configuration 
$\vec{M}^{0}(\vec{r})$ is a N\'eel domain wall (Fig.1b) with center at
$x=0$ and parallel to the $z$-$y$ plane (Fig.1a).
Because $h$ is very small $M^{0}_{z}$ is negligible, 
and $M^{0}_{x}$ and $M^{0}_{y}$ are nearly independent of $z$.  (Due to translation invariance
$\vec{M}$ does not depend on $y$.)  Thus the most general N\'eel domain wall is
\begin{equation}
\label{m_intro}
\vec{M}^{0}(\vec{r})={M_s}\hat{n}={M_s}(u(x),v(x),0) \mbox{,}
\end{equation}
where 
$u(x)^2+v(x)^2=1$, subject to $u(\pm w)=0$, $v(\pm w)=\mp 1$,
with $u(-x)=u(x)$ and $v(-x)=-v(x)$. 
For $\delta \ll w$, near the center of the domain wall, 
$u(x)\approx\sech(x/\delta)$ and $v(x)\approx-\tanh(x/\delta)$. 

Above the critical thickness $h_{c}$ 
the stable configuration is the asymmetric Bloch wall (Fig.1c),  
whose magnetization $\vec{M}^{B}(\vec{r})$ lacks the $x$ and $z$ reflection symmetries, 
but preserves inversion symmetry $(x,z)\rightarrow (-x,-z)$, 
$M^{B}_{x,z}(-x,-z)=M^{B}_{x,z}(x,z)$, $M^{B}_{y}(-x,-z)=-M^{B}_{y}(x,z)$.

The critical mode responsible for the N\'eel-to-asymmetric-Bloch-wall transition 
has magnetization $\vec{\mu}(x,z)=\vec{M}^{B}(x,z)-\vec{M}^{0}(x,z)$ and 
is symmetric under reflections of $x$ and $z$, so
\begin{eqnarray}
&\mu_x(-x,z)=-\mu_x(x,z); \qquad  &\mu_{x,y}(x,-z)=-\mu_{x,y}(x,z) \nonumber \\
&\mu_{y,z}(-x,z)=\mu_{y,z}(x,z); \qquad  &\mu_z(x,-z)=\mu_z(x,z) \mbox{.}
\end{eqnarray}

Since the amplitude of the critical mode is small, its magnetization 
is orthogonal to the static magnetization and can be written as
\begin{equation}
\label{mu}
\vec{\mu}=M_s (\lambda(x,z) v(x),-\lambda(x,z) u(x), \zeta(x,z)),
\end{equation} 
where $\lambda(-x,z)=\lambda(x,z)$, $\lambda(x,-z)=-\lambda(x,z)$, $\zeta(x,z)=\zeta(x,z)$ and $\zeta(x,-z)=\zeta(x,z)$. 

Expanding in small $z$ subject to the above symmetry conditions, with $\tilde{z}=2z/h$, we use the approximations
\begin{equation}
\label{approx}
\lambda(x,z)=\left(\lambda_1 \tilde{z} + \lambda_3 \tilde{z}^3\right)f(x), \quad
\zeta(x,z)=\left(\zeta_0 + \zeta_2 \tilde{z}^2 \right) g(x),
\end{equation}
where $\lambda_0$, $\lambda_2$, $\zeta_1$, and $\zeta_3$ are constants to be determined.

The micromagnetic calculations show that the critical mode is localized near the center 
of the strip and that it shifts the N\'eel domain wall core parallel to $x$.
Thus $\mu_x(x,z)\sim \frac{d}{dx}M^{0}_x(x)$ and 
$\mu_y(x,z)\sim \frac{d}{dx}M^{0}_y(x)$.
By comparison with the fits (\ref{mode_x_magn_approx}-\ref{mode_z_magn_approx}) to the micromagnetic calculations, a good set of approximations is
\begin{equation}
\label{trial_functions}
f(x)= \sech\left(x/\delta\right),\,\,
g(x)= \sech\left(x/\delta\right) \left[2 \sech\left(x/\delta\right) - 1\right] .
\end{equation}

The mode energy now takes the form 
\begin{equation}
\label{hamiltonian:2d:mode}
W_{m}[\hat{n},\vec{\mu}]
=W_{ex}[\hat{n},\vec{\mu}] + W_{dd}[\hat{n},\vec{\mu}]; 
\end{equation}
where $W_{ex}$ and $W_{dd}$ are the exchange and dipole-dipole contributions.
Although the mode is localized in the central part of the domain wall, 
the exchange contribution to its energy is negligible, so\cite{hubert:2001} 
\begin{equation}\label{eq:energy}
W_{m}[\hat{n},\vec{\mu}]=\int d\vec{r}_{1}d\vec{r}_{2}
(\mu(\vec{r}_{1})\vec{\partial}_{r_{1}})
(\mu(\vec{r}_{2})\vec{\partial}_{r_{2}})
\frac{1}{|\vec{r}_{1}-\vec{r}_{2}|}.
\end{equation}

We now expand the free energy functional using 
the trial solution (\ref{mu}-\ref{trial_functions}).  The energy of the mode is calculated
to second order in the $\lambda_i$, $\zeta_j$, and by stability has no first-order terms.
With the row vector $Z=\left(\lambda_1,\lambda_3,\zeta_0,\zeta_2\right)$ and its 
column (transpose) vector $Z^T$, we have
\begin{equation}
W_{m}[\hat{n},\vec{\mu}]=Z \hat{W}_{mode}Z^{T} ,
\end{equation}
where $\hat{W}_{mode}$ is a $4\times 4$ matrix 
whose elements depend on the system parameters.

At the critical thickness $h_{c}$ one of the eigenvalues of $\hat{W}_{mode}$ goes to zero, so its determinant goes to zero:
\begin{equation}
\label{det_eq}
Det\,\hat{W}_{mode}=0.
\end{equation}

For small exchange length $l_{ex}$, as we have here, $\hat{W}_{ex}\ll \hat{W}_{dd}$.  The 
term $\hat{W}_{dd}$ is a function only of $\delta / h$.  
Solving (\ref{det_eq}) 
yields $h_c\approx 0.8 \delta_{c}$ (to about 10\%).    (Here $\delta=\delta_{c}$ is evaluated at the transition.)  This is in good agreement with the numerics, where for different $w$ and $l_{ex}$ we find $0.74 < h_c/\delta_{c} < 0.78$.  The unnormalized critical vector is 
\begin{equation}
Z_{c}=(0.19, 0.70, 0.66, 0.17).
\end{equation}
This yields the form of $\vec{\mu}(x,z)$; its amplitude 
is determined by a higher order expansion in the energy. 

The smallest eigenvalue $\lambda_{0} $ of the matrix $\hat{W}_{mode}$ goes 
to zero linearly when $h$ approaches $h_{c}$. Because the rest of the modes 
have finite energy a standard result of Landau-Lifshitz equations is that 
the frequency of the mode $\omega^{2}\sim \lambda_{0}\sim h_{c}-h$, 
so that the frequency goes to zero at the critical thickness as 
a square root $\omega \sim \sqrt{h_{c}-h}$.

The expression (7) gives only the first order approximation to the
transition mode $\vec{\mu}(x,z)$. The second order correction 
$\vec{\chi}(x,z)$ can be obtained from $|\vec{M}^{B}(x,z)|=
|\vec{M}^{0}(x,z)+\vec{\mu}(x,z)+\vec{\chi}(x,z)|=M_s$ and
$\vec{\mu}(x,z)\perp \vec{\chi}(x,z)$. Using this correction, 
one can calculate the free energy near the transition:
\begin{equation}
W_{mode}(\eta)=\eta^2 0.3 M_s^2 L h^2 \left(\frac{\delta}{h} - \frac{\delta_c}{h_c}\right)
+0.008 \eta^4 M_s^2 L h^2 ,
\end{equation}
where $\eta$ is the amplitude of the critical mode.

Because there is no $\eta^3$ term, the relation between
$W_{mode}(h)$ and the mode frequency $\omega$ is: 
\[\sqrt{2}\left.\frac{\partial \omega}{\partial h}\right|_{h=h_c-0}=
-\left.\frac{\partial W_{mode}}{\partial h}\right|_{h=h_c+0}\,.\]

{\bf Summary and Discussion.} Experimental study of the transition by measurement of the magnetization and even the critical mode would be desirable.  Because the symmetry under $-z$ to $+z$ of the critical mode changes at the transition, from in-phase to out-of-phase, there are drastic symmetry changes in the absorption.  Thus, whereas for $h<h_{c}$ (N\'eel wall) the mode strongly couples to a uniform out-of-plane rf field, for $h>h_{c}$ (asymmetric Bloch wall) it becomes anti-symmetric with respect to the z axis and cannot couple to this same rf field.

To summarize, we have found that the N\'eel to asymmetric Bloch wall transition in dipole-dipole coupled thin magnetic films of infinite length along one direction, with relatively large width $2w$, is a smooth function of thickness $h$.  The normal mode frequencies reveal that at a critical thickness $h_{c}$ the frequency of the transition mode, where the nature of the domain wall changes, goes to zero as $\sqrt{h_{c}-h}$.  To high accuracy, both static and dynamic calculations indicate that this transition is continuous, as also supported by analytical studies. 

\begin{acknowledgments}
We gratefully acknowledge the support of the Department of Energy through Grant DE-FG02-06ER46278.
\end{acknowledgments}


\begin{thebibliography}{0}

\bibitem{bitter:1903} F. Bitter, 
Phys. Rev. 38, 1903-1905 (1931). 

\bibitem{bloch:1932} F. S. Bloch, 
Zeitschrift f\"ur Physik 74, 295-335 (1932). 

\bibitem{landau:1935} L. D. Landau, and E. M. Lifshitz, 
{Physik. Z. der Sowjetunion} {8}, {153--169} (1935). 

\bibitem{neel:1955} L. N\'eel, 
Comptes Rendus Hebdomadaires des S\'eances de l'Acad\'emie des Sciences 421, 533-536 (1955). 

\bibitem{middlehoek:1966} S. Middlehoek, 
IBM J. Res. 10, 351-354 (1966). 

\bibitem{scheinfein:1989} M. R. Scheinfein, J. Unguris, R. J. Celotta, and D. T. Pierce, 
Phys. Rev. Lett. 63, 668-671 (1989). 

\bibitem{scheinfein:1991} M. R. Scheinfein, J. Unguris, J. L. Blue, K. J. Coakley, D. T. Pierce, R. J. Celotta, and P. J. Ryan, 
Phys. Rev. B 43, 3395--3422 (1991). 

\bibitem{melcher:2003} C. Melcher, 
Arch. Rat. Mech. Anal. 168, 83--113 (2003). 

\bibitem{RivkinRomanov-Neel:2007} K. Rivkin, K. Romanov, Y. Adamov, Ar. Abanov, W. Saslow, and V. Pokrovsky, Phys. Rev. B (submitted to Phys. Rev. B). 

\bibitem{chang:2007} C. C. Chang, Y. C. Chang, I. C. Lo, and J. C. Wu, 
J. Magn. Magn. Mat., 310, 2612--2614 (2007). 

\bibitem{redjdal:2002} M. Redjdal, J. Giusti, M. F. Ruane, and F. B. Humphrey, 
{J. Appl. Phys.}, 91, {7547--7549} (2002). 

\bibitem{hubert:1975} A. Hubert, 
IEEE Transactions on Magnetics 11, 1285--1290 (1975). 

\bibitem{metlov:2005} K. Metlov, 
J. Low Temp. Phys. 139, 207-219 (2005).

\bibitem{trunk:2001} T. Trunk, M. Redjdal, A. Kakay, M. F. Raune, and F. B. Humphrey,
J. Appl. Phys. 89, {7606--7608} (2001). 

\bibitem{kakay:2005} A. Kakay, {\it Numerical investigations of micromagnetic structures}, 
Research Institute for Solid State Physics and Optics, Hungarian Academy of Sciences (2005). 

\bibitem{rivkin:2004} K. Rivkin, A. Heifetz, P. R. Sievert, and J. B. Ketterson, 
Phys. Rev. B 70, {184410} (2004). 



\bibitem{hubert:2001} A. Hubert, and R. Schafer, {\it Magnetic Domains: The Analysis of Magnetic Microstructures}, Springer (2001), p.148. Eq.(3.61). 

\end{thebibliography}


\end{document}